# Superconducting Dome and Crossover to an Insulating State in $[Tl_4]Tl_{1-x}Sn_xTe_3$

K.E. Arpino,[1,2] B.D. Wasser,[3] and T.M. McQueen[1,2,3,a)]

[1]*Department of Chemistry, The Johns Hopkins University, Baltimore, Maryland 21218, USA*

[2]*Institute for Quantum Matter, Department of Physics and Astronomy, The Johns Hopkins University, Baltimore, Maryland 21218, USA*

[3]*Department of Materials Science and Engineering, The Johns Hopkins University, Baltimore, Maryland 21218, USA*

The structural, superconducting, and electronic phase diagram of $[Tl_4]Tl_{1-x}Sn_xTe_3$ is reported. Magnetization and specific heat measurements show bulk superconductivity exists for $0 \leq x \leq 0.4$. Resistivity measurements indicate a crossover from a metallic state at $x = 0$ to a doped insulator at $x = 1$. Universally, there is a large non-Debye specific heat contribution, characterized by an Einstein temperature of $\theta_E \approx 35$ K. Density functional theory calculations predict $x = 0$ to be a topological metal, while $x = 1$ is a topological crystalline insulator. The disappearance of superconductivity correlates with the transition between these distinct topological states.

[a)]Author to whom correspondence should be addressed.  Electronic mail:  mcqueen@jhu.edu.

With the discovery of topological insulators, the importance of spin-orbit coupling in driving electronic phenomena has moved to the forefront of condensed matter physics[1,2,3]. Breaking time-reversal or inversion symmetry in a crystalline material can lift Kramer's degeneracy of spins, resulting chiral locking of spin and momentum displayed by topological features such as Dirac cones in the surface states of topological insulators and, more recently, in the three-dimensional cone of Dirac semi-metals[4,5,6,7,8,9]. Topological materials are now known to offer an exciting array of physical phenomenon, such as Fermi arc surface states and Majorana fermions, with a wealth of potential applications[10,11,12,13]. Many of these phenomena are predicted to occur at the convergence of topological states and other emergent states of matter. More recently, several candidate materials combining topological surface states and bulk superconductivity, including $Cu_xBi_2Se_3$[14], $Au_2Pb$[15] and $[Tl_4]TlTe_3$ ($Tl_5Te_3$)[16], have been experimentally proposed.

Topological phase transitions and their attendant effects, such as fractional charge and spontaneous mass acquisition, have been previously examined in $(Bi_{1-x}In_x)_2Se_3$[17] and $BiTl(S_{1-x}Se_x)$[18,19]. The $[Tl_4]Tl_{1-x}Sn_xTe_3$ series is also predicted to exhibit a doping-induced topological phase transition: calculations have indicated that $[Tl_4]TlTe_3$, but not $[Tl_4]SnTe_3$, harbors a non-trivial $Z_2$ topological invariant when appropriately electron doped[16]. This series has the additional interest of harboring bulk superconductivity, as end-member $[Tl_4]TlTe_3$ superconducts below $T_c = 2.4$ K while $[Tl_4]SnTe_3$ does not.

Here we report the structural and electronic phase diagram of $[Tl_4]Tl_{1-x}Sn_xTe_3$ over the entire solid solution range $0 \leq x \leq 1$. Samples were prepared as previously reported, yielding cleavable crystalline boules[16]. Magnetization and specific heat measurements show a superconducting dome upon substitution of Sn for Tl, with a maximum $T_c = 2.73(4)$ K at $x = 0.1$. From resistivity measurements, this superconducting dome is proximal to a crossover to an insulating state around $x \approx 0.5$. Throughout the entire solid solution, there is a large non-Debye contribution to the lattice specific heat. Band structure calculations consistent with these experimental observations suggest that the system is transformed from a $Z_2$ topological metal at $x = 0$ to a topological crystalline insulator at $x = 1$.

The structure of $[Tl_4]Tl_{1-x}Sn_xTe_3$ (inset of **Figure 1**) can most conveniently be viewed as a derivative of perovskite ($ABO_3$): a three-dimensional network of corner-sharing $(Tl_{1-x}Sn_x)Te_6$ octahedra houses interconnected $[Tl_4]$ tetramers in the cavities between octahedra[16,20]. The structure is tetragonal due to alternating octahedral rotations along the $c$ axis (Glazer notation $a^0a^0c^-$)[21] which are necessary to accommodate the $[Tl_4]$ units[16,22]. **Figure 1** shows the evolution of lattice parameters as a function of Sn content. There has been some question about the doping limit and location, with previous reports variously claiming the maximum possible Sn content as ranging from 0.8 to 1.1[20,22,23]. Our data clearly show a continuous evolution of lattice parameters up to $x = 1.0$, followed by a plateau when $x > 1$, demonstrating that substitution of up to one molar equivalent of Sn for Tl is possible. The limit of $x = 1$ is consistent with charge counting: assuming dianionic Te, the formal charges are $[Tl_4]^{4+}Tl^{2+}Te_3^{2-}$; divalent $Sn^{2+}$ is then able to substitute for all divalent, but not monovalent, Tl, i.e, up to one molar equivalent.

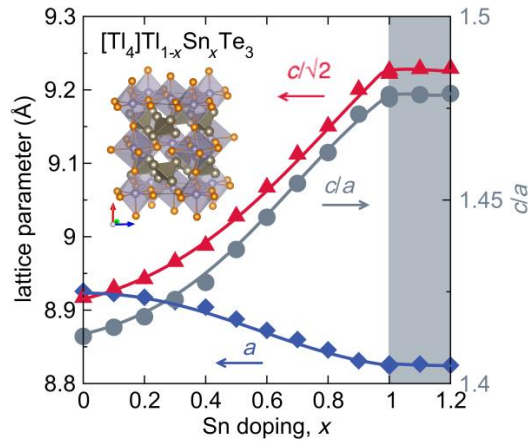

**Figure 1.** The lattice parameters of $[Tl_4]Tl_{1-x}Sn_xTe_3$ show continuous, but non-linear, evolution with increasing Sn content, with a solid solubility limit at $x = 1$. The $c$ lattice parameters (red triangles) divided by a factor of $\sqrt{2}$ (a measure of pervoskite pseudo-cubicity) and the $a$ lattice parameters (blue diamonds) are plotted on the leftmost axis, while the $c/a$ ratio (gray circles) is plotted on the right axis. Inset[24]: The material is a perovskite analogue, with $(Tl_{1-x}Sn_x)Te_6$ octahedra and interstitial $[Tl_4]$ tetramers. The octahedra are significantly rotated in an $a^0a^0c^-$ fashion[16,21,22].

The nominal isovalent replacement of $Tl^{2+}$ by $Sn^{2+}$ dramatically modulates the physical properties. **Figure 2**(a) shows low field ac magnetization measurements for $0 \leq x \leq 0.4$, measured using a quantum design physical properties measurement system (QD-PPMS). Consistent with previous reports, the $x = 0$ endmember is superconducting with a $T_c = 2.36(6)$ K[16,25]. $T_c$ increases initially with Sn substitution, to a maximum of $T_c = 2.73(4)$ K at $x = 0.1$, before falling for $x > 0.4$. The bulk nature of the superconductivity is confirmed by the presence of a lambda anomaly in the low-temperature electronic specific heat [**Figure 2**(b)], measured using the semi-adiabatic pulse technique as implemented in a QD-PPMS. The specific heat jump at $T_c$, $\Delta C_{el}/(\gamma T_c) = 1.6$, changes little as a function of doping and is close to the weak coupling Bardeen-Cooper-Schrieffer (BCS) value of 1.43[26]. Only traces of superconductivity remain for $x > 0.4$.

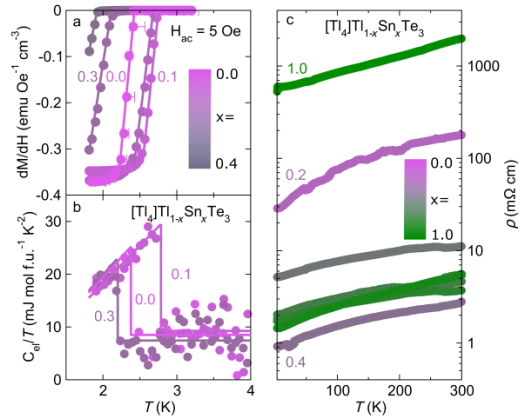

**Figure 2.** The superconductivity of compounds $0 \leq x \leq 0.4$ is shown in the negative ac magnetic susceptibility response (a) and low-temperature heat capacity measurements (b). Equal entropy constructions on the electronic heat capacity illustrate the bulk nature of the superconductivity. (c) Resistivity measurements of the $[Tl_4]Tl_{1-x}Sn_xTe_3$ series show the a non-monotonic change in electrical conductivity for $T \geq 3K$ as a function of Sn content. Data from the fully Sn-doped end member is consistent with a semiconductor, while that of the majority of the series indicates semi-metallic or metallic states.

Resistivity measurements [**Figure 2**(c)] collected using a QD-PPMS with a four-probe geometry and a constant excitation current show a non-monotonic change in electrical conductivity as a function of Sn content. The magnitude of the resistivity over most of the series is consistent with a semi-metallic or metallic state, while the highest doping ($x = 1$) is consistent with the behavior expected for a doped semiconductor. This transition from metallic towards insulating behavior upon Sn doping is in agreement with previous measurements[23].There is a concomitant change in the shape of the temperature dependence that also suggests a change in the dominant scattering mechanism across the series.

Accompanying these changes in physical properties are changes in the structure. The lattice parameters, **Figure 1**, show a monotonic trend as a function of $x$, but with pronounced deviations from the linear behavior expected from Vegard's law[27]. Rietveld analysis of powder diffraction data was used to elucidate the structural changes with composition. The non-linearity of the $a$ lattice parameter arises due to a combination of two structural effects in the $a$-$b$ plane highlighted in **Figure 3**. First, there is a linear reduction of the in-plane octahedral metal-tellurium distance, as expected given the smaller size of Sn versus Tl. Second, there is a concomitant linear reduction of the degree of the $c$-axis octahedral rotations, which likely arises from the need to maintain optimal bonding geometry to the [Tl$_4$] subunits. This linear decrease in rotation angle corresponds to a non-linear-decease in the distance between in-plane $B$-sites, resulting in the non-linearity of the $a$ lattice parameter

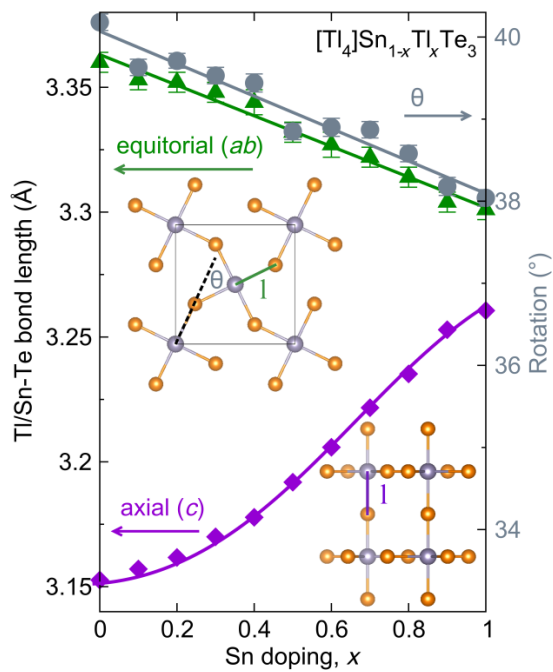

**Figure 3.** The octahedral bond lengths are shown. Those bond distances between the metal center and tellurium in the *a-b* plane (green traingles) exhibit a linear decrease, while those between the metal center and axial tellurium along the *c* axis direction (purple diamonds) increase non-linearly. The octahedral rotation from straight is plotted on rightmost axis, showing a linear decrease.

More interesting is the non-linear, and positive, change of the *c* axis upon Sn substitution: in the reported *I*4/*mcm* unit cell, there are no internal degrees of freedom along this axis, and the distance is set by the size of the *B*-site metal cation and the axial (out-of-plane) octahedral metal-tellurium bond lengths. Thus from cation size, a linear decrease is expected upon Sn substitution instead of the observed non-linear increase. **Figure 3** shows the axial (out-of-plane) octahedral metal-tellurium distance increases non-linearly with Sn substitution, which mirrors the change in *c* as required by the Wyckoff positions of the involved atoms. Given the smaller size of Sn relative to Tl and their isovalency, this increase in bond distance is unexpected. This effect is not an artefact of an incorrect choice of symmetry: refining in lower-symmetry space group *I*4/*m*, which allows for movement of the axial Te along the *c* axis, does not offer a significant change in the bond length. Instead it likely reflects the differences in the covalency between Tl-Te and Sn-Te bonding.

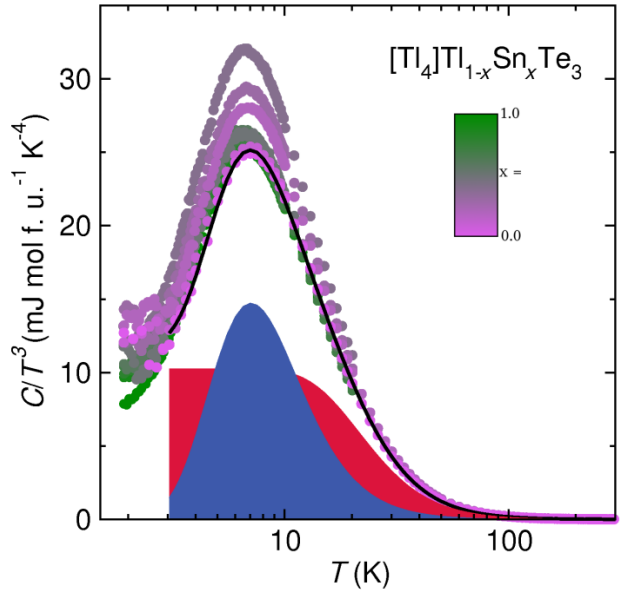

**Figure 4.** The lattice contributions to the heat capacity only vary slightly as a function of Sn content. However, there is a substantial non-Debye contribution centered at $T \approx 7$ K, which is well described as an Einstein oscillator. The overall fit (black line) for $x = 0$, with corresponding Debye (red) and Einstein (blue) components, is shown.

In addition to the structural changes, a substantial non-Debye lattice contribution was observed in the specific heat of all samples; **Figure 4** shows the specific heat for the series of samples, plotted as $C/T^3$ vs. $\ln T$ [28]. The specific heat data are well described above $T_c$ as one Debye mode, one Einstein mode, and an electronic heat contribution given by the Sommerfeld term [29]. The large Einstein contribution at low temperature obscures the precise behavior of the Sommerfeld coefficient with composition, but the electronic contribution appears to correlate with the superconducting $T_c$, with both showing a maximum at x = 0.1. The Debye component is characterized by a Debye temperature of $\theta_D =$ 97 K - 117 K and 6.5 - 6.9 oscillators per formula unit. An increase in the Debye temperature upon Sn substitution is expected due to the lighter mass of Sn relative to Tl. The Einstein mode has parameters Einstein temperature of $\theta_E$ = 37 K - 33 K and 1.1-1.5 oscillators per formula unit, with a maximal contribution for $x = 0.4$, and likely originates from a low-lying optic phonon mode. Such modes are commonly observed in ferroelectrics and materials near structural instabilities[30,31]. As $Tl^{2+}$ is a negative-U ion, favoring charge disproportionation, and $Sn^{2+}$ can be lone-pair active, it is alluring to ascribe this Einstein contribution to a propensity toward charge order ($x = 0$) or off-centering of the B-site cation ($x$

= 1). However, such an assignment is unlikely in this case since the total Einstein contribution shows very little dependence on composition: it would be quite a coincidence for two different origins to produce a low-lying optic mode at the same energy. Instead, the fact that this contribution is consistent for all $0 \leq x \leq 1$ suggests the origin of instead lies within the [Tl$_4$] framework, the only units not directly disturbed by the Sn substitution at the $B$-site.

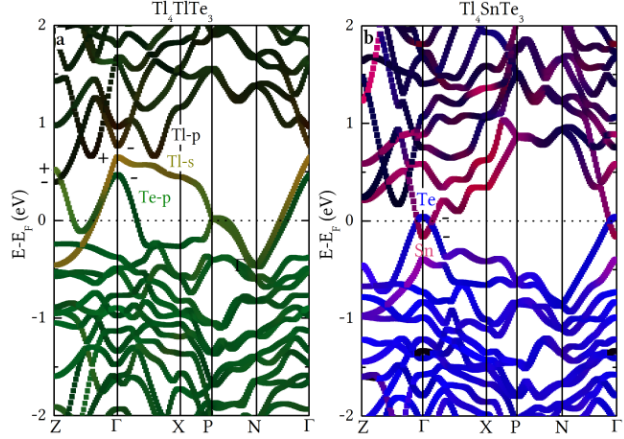

**Figure 5.** DFT band structures including spin orbit coupling for (a) [Tl$_4$]TlTe$_3$, U = -5 eV, and (b) [Tl$_4$]SnTe$_3$. Shading in (a) indicates contributions from various atomic orbital contributions, while shading in (b) indicates contributions from the stated atoms. The parity of states near the Fermi level relevant to determining the topological class are given by +/- symbols. [Tl$_4$]TlTe$_3$ is found to have a $Z_2$ invariant of -1 and is a topological metal, while [Tl$_4$]SnTe$_3$ is topologically trivial but has an inversion of Sn and Te derived states at Γ that produce a topological crystalline insulator state.

To gain insight into the origins of the changes across the series, density functional theory (DFT) calculations were performed on the two end-members using LDA+U as implemented in the full potential linearized augmented plane wave plus local orbitals (FP-LAPW-LO) code elk[32] and a 6x6x4 k-point mesh; the resulting band structures are plotted in **Figure 5**(a) ($x = 0$) and **Figure 5**(b) ($x = 1$). In agreement with previous calculations that did not include the negative U effect of Tl$^{2+}$,[16] [Tl$_4$]TlTe$_3$ is found to contain both electron and hole pockets, and is best described as a metal. Due to the $I4/mcm$ symmetry, to calculate the $Z_2$ topological invariant, it is only necessary to consider the states at the Γ and Z time reversal invariant points in the Brillouin zone. Multiplying the parities of all *occupied* states at Z and Γ for [Tl$_4$]TlTe$_3$ give a total parity of $Z_2 = -1$; i.e., Tl$_5$Te$_3$ is topologically nontrivial. This is due to a strong covalency between Tl *s* and Te *p* states that pushes a negative-parity Te-*p* derived band above $E_F$ at the Γ point. In contrast, SnTl$_4$Te$_3$ is topologically trivial: there is a band inversion at the Γ point, but it is between two bands of negative parity and does not affect the $Z_2$ invariant. However, the inversion at Γ does move Sn *p* states into the valence band and Te *p* states into the conduction band. The result is analogous to the band inversion in SnTe, and is expected to give rise to topological crystalline insulator states[4].

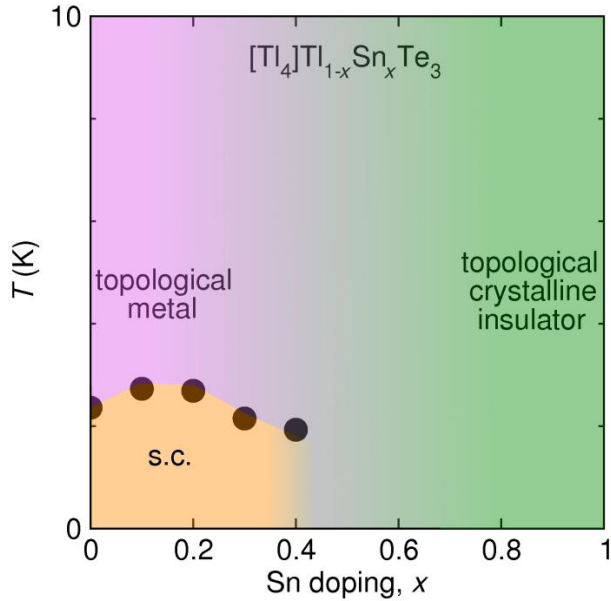

**Figure 6.** Phase diagram of [Tl$_4$]Tl$_{1-x}$Sn$_x$Te$_3$ showing the superconducting dome proximal to a transition from a topological metal to topological crystalline insulator.

A summary of our experimental and computational results is presented in **Figure 6**. Substitution of Sn for Tl in [Tl$_4$]TlTe$_3$ results in the rise and fall of superconductivity in proximity to a crossover between a topological metal and a topological crystalline insulator. It is inappropriate to speculate whether the disappearance of superconductivity is tied to the change in topological class in this fascinating set of materials. Further work is required to elucidate the precise origin of the large non-Debye lattice contribution to the specific heat observed across the series. We expect these results to motivate significant additional studies into materials that couple topological surface states to other degrees of freedom.

This work was supported by the David and Lucile Packard Foundation. This research benefited from structural studies performed using beamlines 11-BM-B and 11-ID-B of the Advanced Photon Source, a U.S. Department of Energy (DOE) Office of Science User Facility operated for the DOE Office of Science by Argonne National Laboratory under Contract No. DE-AC02-06CH11357. We thank David Wallace and John Sheckelton for technical assistance.